\documentclass[eqno,twoside,11pt]{article}

\usepackage[dvips]{graphicx}
\usepackage{epsfig}
\usepackage{amsfonts}
\usepackage{amssymb}
\usepackage{epsfig}
\usepackage{subfigure}
\usepackage{amssymb}
\def\bbbz{{\mathchoice {\hbox{$\sf\textstyle Z\kern-0.4em Z$}}
{\hbox{$\sf\textstyle Z\kern-0.4em Z$}}
{\hbox{$\sf\scriptstyle Z\kern-0.3em Z$}} {\hbox{$\sf\scriptscriptstyle
Z\kern-0.2em Z$}}}}
\def\bbbI{{\rm 1\kern-0.25em I}}

\begin{document}

\newlength{\au}
\settowidth{\au}{\normalsize \rm 01-815~Warszawa,~Poland aa}

\title{Zipf's law and phase transition}
\date{}
\maketitle
\begin{center}
{{\large \textbf{ $\mathrm{Krystyna\; Lukierska-Walasek}$}, }}\\
\bigskip
{Faculty of Mathematics and Natural Sciences\\
Cardina{\l} Stefan Wyszy\'nski University, ul.~Dewajtis 5 , 01-815 Warsaw, Poland}\\
e-mail: k.lukierska@uksw.edu.pl \\
\bigskip
{{\large \textbf{$\mathrm{Krzysztof\;Topolski}$}}}\\
\bigskip
 {Institute of Mathematics, Wroc\l aw University\\ Pl.~Grunwaldzki 2/4, 50-384 Wroc\l aw, Poland}\\
e-mail: topolski@math.uni.wroc.pl
\end{center}
\begin{abstract}
 We describe the link between  the Zipf law and  statistical distributions for the Fortuin-Kasteleyn clusters in  Ising
as well as Potts models. From these results it is seen that Zipf's law can be a criterion of a phase transition, but it does not determine its order . We present the corresponding histograms for fixed domain configurations.
\end{abstract}




\section{Introduction}
\label{I}
Originally the Zipf law was used for analyzing the hierarchy of word's occurance in a language, i.e. the relative population of words ranking from these used most frequently to the ones used less frequently \cite{Zip}. Existence of very similar linear hierarchy distributions was found in other domains of science, for example the Zipf law has appeared in the description of distributions of city populations  \cite{Bat}, distributions of turnovers of Europe's largest companies \cite{Bou}, in linguistic features of noncoding DNA sequences \cite{Mbg}, molecular biology networks \cite{Sne} and in physics in multifragmentation of atomic nuclea in nuclear reactions \cite{Ma}, \cite{Ma1} or liquid crystal \cite{Sci}.
The statistics of domains was numerically investigated for percolation \cite{Sta}, \cite{Wat}, for Ising and Potts models \cite{Car}, \cite{Jan}, \cite{Klt}, \cite{Klt2}.  Various formulation of cluster theories one can find in papers dealing with Fisher droplets \cite{Fis} or percolation \cite{Sta}.
The Zipf law states that the mass $m$ of the largest, second largest, ..., $k-$largest clusters decreases according to their rank, $k=1,2,3,..,n$, as
\begin {equation}
x(k) \sim\frac{1}{k^\lambda},\quad \mbox{where}\,\, \lambda \sim 1.
\end{equation}
 We wish to underline that Zipf's law is a special case of the more general  Zipf-Mandelbrot law \cite{Man}
\begin {equation}
x(k) = C/( k + \alpha)^{-\lambda},
\end{equation}
where the offset $\alpha$ is an additional constant parameter. The value of $\lambda$ is asymptotically approximated by the function of the critical exponent $\tau$ of power law cluster size distribution\cite{Man} and is given  by formula \cite{Man}
\begin {equation}
\lambda = \frac{1}{\tau-1}.
\end{equation}
Zipf's law is a consequence of power law cluster size distribution with the exponent $\tau=2$.
There is a connection of the critical exponent $\tau$ with  critical exponents of scaling theory, similarly as in percolation \cite{Sta}.

We study the size distributions of the Fortuin-Kasteleyn (FK) clusters obtained by the Monte Carlo simulations in a pair of models - two-dimensional Ising and Potts models- considered as well in our recent publications \cite{Klt}, \cite{Klt2}.
In our case the measure of size is the number of spins in the cluster,defining the mass of the cluster. For the FK spin clusters, near the critical point, the probability density distribution of the number of clusters with mass $x$ has the asymptotic form
\begin {equation}
\rho(x)\sim x^{\tau}\exp[{-\theta x}].
\end{equation}
The first factor  characterized by exponent $\tau=\frac{d}{D}+1$ is the entropy factor in where $d$ is the dimension of the system and $D$ describes its fractal dimension of the system. The second factor is the Boltzman weight, which suppress large clusters when parameter $\theta$ is finite.
When parameter $\theta$ tends to zero we can use the approximation $\theta\sim |T-T_c|^{\frac{1}{\sigma}}$ which defines a critical exponent $\sigma$.

We investigate the statistics of the domain masses for  Ising and Potts models in the critical region as well beyond it. Connection of the exponent $\tau$ of the cluster size distribution with critical exponents of scaling theory causes that Zipf's law can be treated as providing some criterion of a phase transition.\\
 We  notice that the power law distributions are not the only forms of broad distribution. Zipf's law suggests also modeling by distributions of hyperbolic type, as in linguistics \cite{Har} and economy \cite{Bib}. Generalized inverse Gaussian distributions play an important role in the theory of generalized hyperbolic distributions. Their right hand tail behaviour spans a range of formulas from exponential decrease to a Pareto tail. They are much slower than in the case of normal distribution and therefore suitable to describe model phenomena, where numerically large values are more probable.
\section{Application of Zipf's law to Ising and Potts models} \label{II}
The Potts model \cite{Pot} is a generalization of the Ising model \cite{Isi} to spins with more than two  components; for a detailed review of the Potts model see \cite{Wu}.

The  Hamiltonian for the $q-$state Potts model \cite{Wu} is the following
\begin{equation}
 H=-\sum\limits_{ij} J_{ij}\delta_{\sigma_i,\sigma_j},
\end{equation}
where $\sigma_{i}\in\{1,2,...,q\},$ $\delta_{x,y}$ is the Kronecker delta
$$\delta _{x,y}=\left\{
\begin{array}{l}
1 \quad \mbox{if \ $x=y$,}\\[2mm]
0 \quad \mbox{otherwise}.
\end{array}\right .
$$
and
$$
J_{ij}=\left\{
\begin{array}{l}
J \quad \mbox{if \ $i,j$ describe neighbour pairs of spins,}\\[2mm]
0 \quad \mbox{in opposite case}.
\end{array}\right .
$$
The case $q=2$ describes  the Ising model.

A number of exact results for  two-dimensional Potts models are known in the infinite volume limit. For example, the phase transition appears at the critical temperature $\,T=T_c\,$
($\,T_c=2J/k_B\,\ln(1+\sqrt{q})\,$). It means that for Ising model $T_c=2.2692$ while for Potts model with $q=3$, $T_c=1.9899$ and with $q=6$, $T_c=1.6152$
It is the second order phase transition for $q\leq 4$ and the first order one for $q\geq 5$, (see  \cite{Wu}).
We use the Fortuin-Kasteleyn random cluster model representation  \cite{Fok}.
Clusters were generating  by  Monte Carlo techniques,
based on the Swendsen-Wang cluster algorithm \cite{Sww} applied to the two-dimensional models
with periodic boundary conditions.
We examine the changes in the cluster size distribution as the system approaches the critical point of the phase transition indicated by the critical temperature.\\
The power law is presented by the Zipf- Mandelbrot law (2)
where $C$ and $\alpha$   are some constants, $\mu$=$\lambda^{-1}$ and $x$ denotes  the value( mass) of the object, $k$ is the rank order of the object. The object with the largest variable value is ranked as the first, the next with  smaller value is ranked as the second and so on. In this way we determine the rank order $k$ of the object.
For $ \alpha = 0 $ and $\mu = 1$ the Zipf-Mandelbrot law  takes the form of  the Zipf law:
\begin{equation}
 x \sim  C/k
\end{equation}
This equation describes a straight line in a double logarithmic plot with the slope equal to $-1$.\\

We can consider the Zipf law  in the Bouchaud notation \cite{Bou}. The distribution of the cluster size appears to be
$[\mu]$-distribution and the indeks $\mu$ is a critical exponent.
Suppose that a set of $N$ $[\mu]$ -variables $ x_{k}$ is ordered as decreasing sequence, then
\begin{equation}
x_{k} \sim x_{0}(N/k)^{1/\mu}
\end{equation}
The formuła (8) means that the largest variable is of the order $x_{0}N^{1/\mu}$, while the smallest is of order $x_{0}$.

In our case $x_k$ is the number of  Ising or Potts spins in the domain with rank $k$, where the greatest cluster, the one with greatest mass, has  rank equal to 1.\\
The Pareto distribution implies the Zipf law  \cite{Tro},\cite{Klt}, but  inverse conclusion is not valid - for instance the Zipf law is satisfied as well in the case of the hyperbolic distributions \cite{Har}, \cite{Bib}.\\
The probability density of the Pareto distributions is defined as follows:
\begin{equation}
\rho(x) = (\mu/x_{0})(x{_0}/x)^{\mu+1}
\end{equation}
for $x > x_0$, where $x_0$ denotes a typical scale, $x$ is the mass and $\mu \sim 1.$

One can define  $\mu$-variable as in \cite{Bou}. The main property of $[\mu]$-variable is that all its momenta $ m_q = < X^q > $ with $q \geq \mu $ are infinite.
\section{Numerical considerations}
 In Ising and Potts models the clusters form sets of nearest neighbour sites occupied by spins with the same orientation.\\
  The $q-$state Potts model has the grand state degeneracy $q$ and after quench from high temperature phase small domains start to grow. There is well established connection between thermal and geometrical phase transitions when the big clusters appear.

\begin{figure}[h]
\begin{center}
\subfigure[Ising]{
\includegraphics[width=5cm]{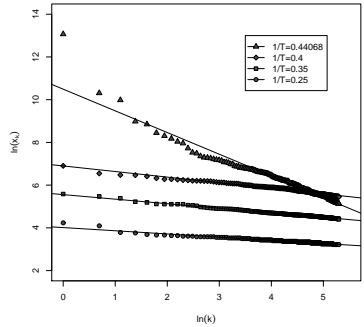}
\label{Ising}
}

\subfigure[3-Potts]{
\includegraphics[width=5cm]{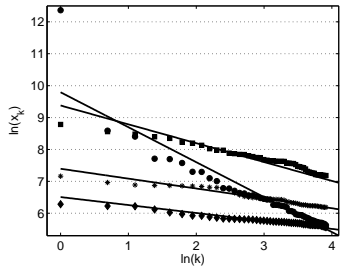}
\label{3-Potts}
}
\subfigure[6-Potts]{
\includegraphics[width=5cm]{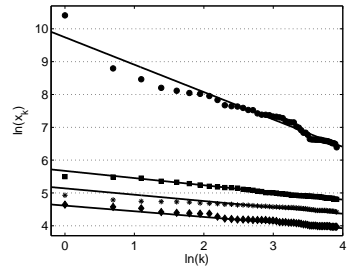}
\label{6-Potts}
}
\caption{\label{Fig1} The log-log distribution of the domain masses $x_k$ versus the rank order index $k$,   for  \ref {Ising} the two-dimensional Ising model for $ T=2.2692$ (the critical temperature) ($\triangle$), $T=2.5000$ ($\Diamond$), $T=2.8571$ ($\Box$) and $T=4.0000$ ($\circ$) for  $L=1000$, \ref{3-Potts} for the 3-state Potts model for $T=2.5000$ ($\Diamond$), $2.2222$ ($\ast$), $2.0408$
($\Box$) and critical $T$ ($\circ$) and \ref{6-Potts} for the 6-state Potts model for  $T=2.0000$ ($\Diamond$), $1.8181$ ($\ast$), $1.6949$
($\Box$) and critical $T$ ($\circ$), for one configuration (see \cite{Klt2}, \cite{Klt})}
\end{center}
\end{figure}
In both models $x_k$ denotes the domain mass of rank $k$.
{\bf Figure \ref{Fig1}} describes  the plot of domain masses versus its rank for the Ising \ref{Ising}, for 3 state Potts  \ref{3-Potts} and 6-state Potts \ref{6-Potts}  models.
The  values of the parameter $\mu$ which is equal to $-{(\tan {\alpha})}^{-1}$, where $\alpha$ is the angle of the slope of  regression line, are significantly smaller than 1 for temperatures different from critical temperature, while for the critical temperature the  value of parameter $\mu$ is approximately equal to 1.
We start from the paramagnetic phase and we decrease the temperature of the system. For Ising model the calculations were done
for  four different temperatures $ T=4.0000$, $T=2.8571$, $T=2.5000$ and $T_{c}= 2.2692 $. The first three absolute values of slopes are significantly smaller than $1$ and equal respectively to $0.149$, $0.213$ and $ 0.261$,  but for the phase transition temperature $T_{c}=0.2692$ the slope is near to $ -1$.\\
We have  similar results for  $3-$state Potts and $6-$state Potts models presented on  Figure 1b and Figure 1c.
 For $3$-statePotts  model the calculations were done for  four different temperatures  $T=2.5000$, $T=2.2222$, $ T=2.0408$ and $T_{c}=1.9900$ respectively with slopes $0.249$, $0.310$, $0.591$ and $1.095$.
 For systems in their critical temperatures the slopes are equal to $-1$ what means that only for this temperatures the Zipf law is satisfied.  We observe that Zipf's law can be used as additional criterion to determine the location of phase transition. Further we can notice that for temperatures higher than critical the domains are ordered along strait lines.\\
  It is interesting, that Zipf law helps to determine location of phase transition without determining its order.
As was noticed by Ma from lattice gas simulations  \cite{Ma}, \cite{Ma1}, it appears that some  questions concerning  the order phase transition still  remain open.

\begin{figure}[h]
\begin{center}
\subfigure[Ising]{
\includegraphics[width=6cm]{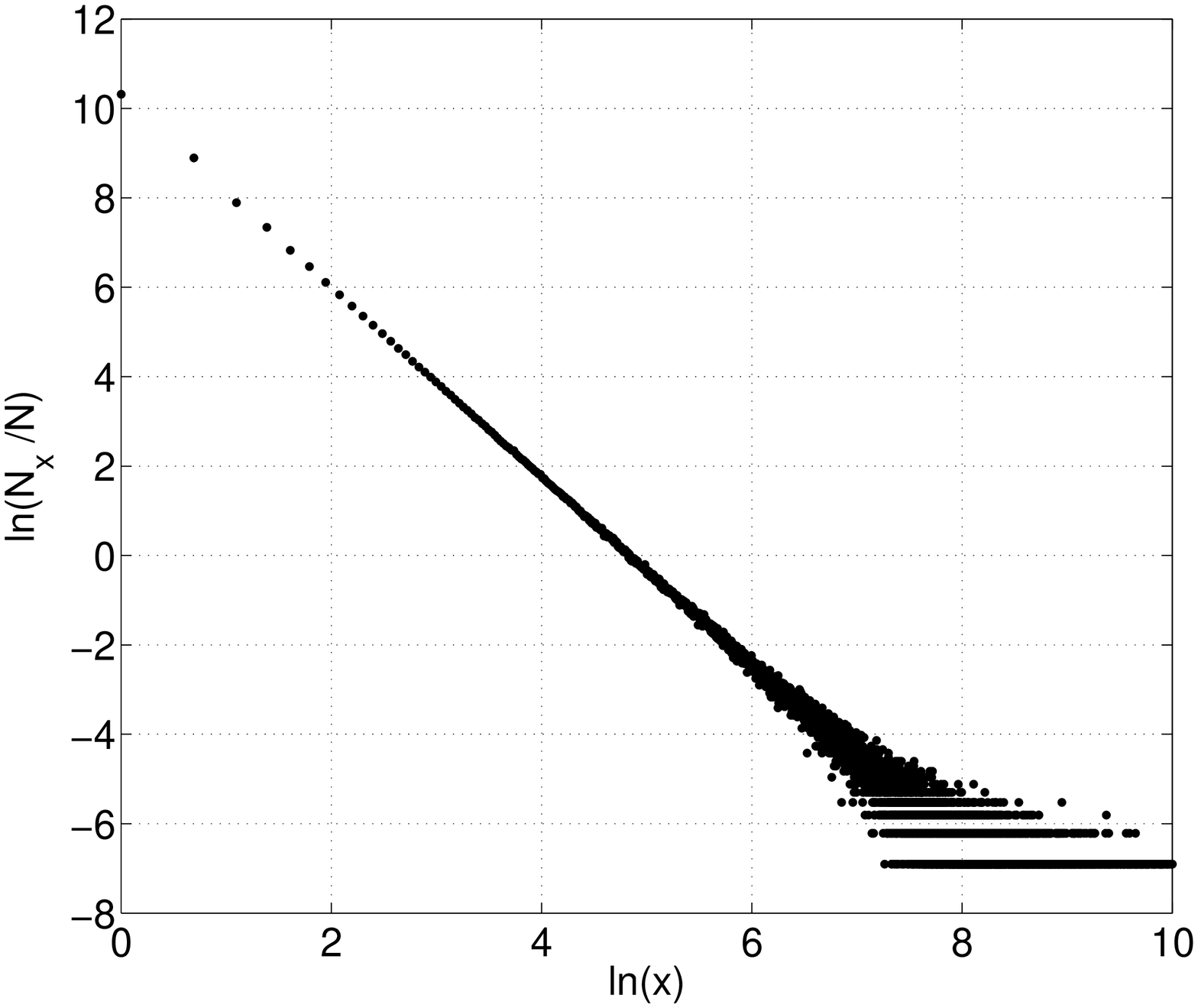}
\label{Ising1}
}

\subfigure[3-Potts]{
\includegraphics[width=5cm]{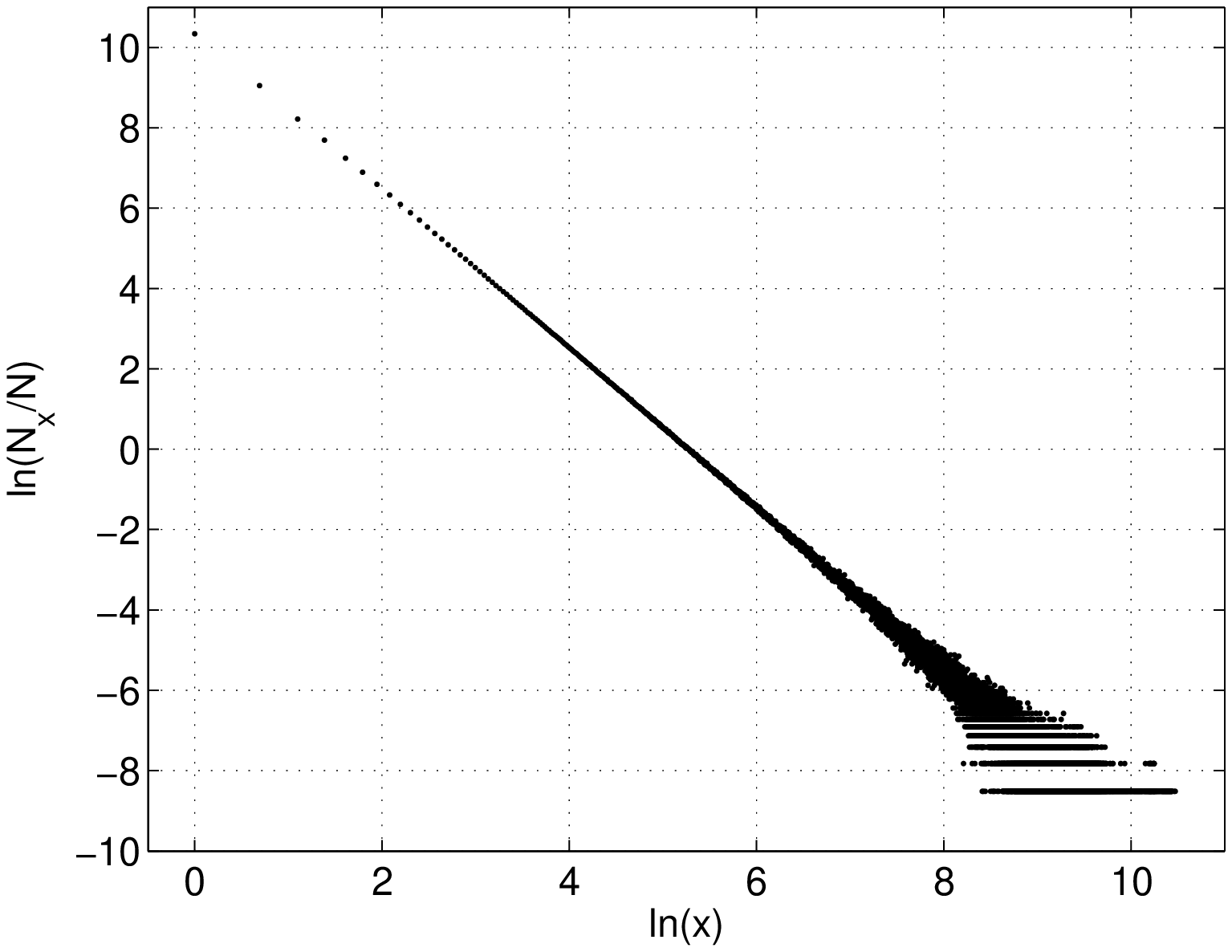}
\label{3-Potts1}
}
\subfigure[6-Potts]{
\includegraphics[width=6.3cm]{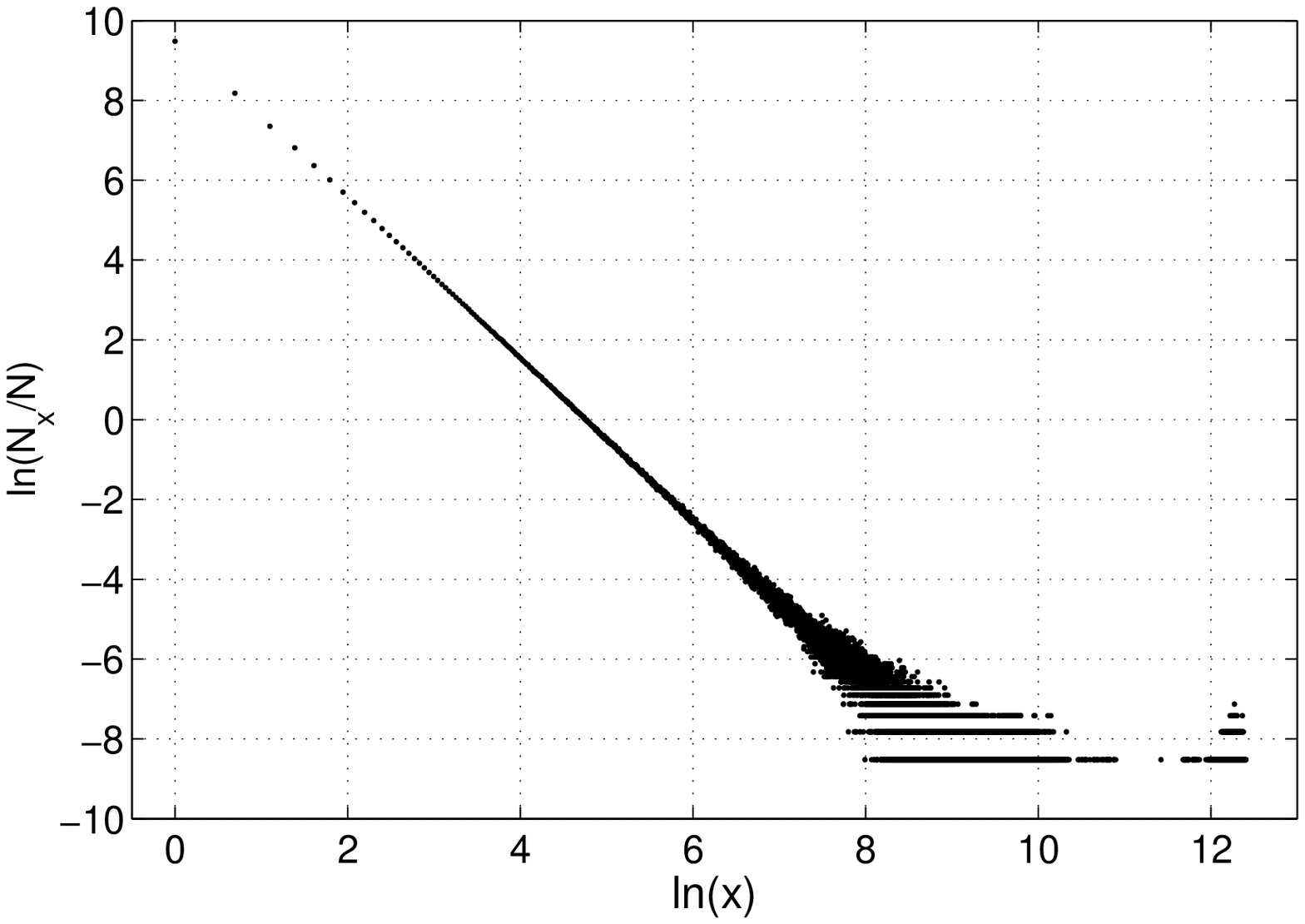}
\label{6-Potts1}
}
\caption{\label{Fig2} The dependence of the  number of domains with mass x normalized by the number of configurations $(N_x/N)$ on the mass(x) in the log-log scale, \ref{Ising1} for the two dimensional Ising model , \ref{3-Potts1} for the 3-state Potts model and \ref{6-Potts1} for the 3-state Potts model, near its criticaltemperatures $T_c$ and  $L=600$.( for 2b and 2c see \cite{Klt})}
\end{center}
\end{figure}
{\bf Figure \ref{Fig2}} presents  the dependence between the number of domain with mass $x$ normalized by number of configurations and the  mass $x$ in log-log scale for Ising model \ref{Ising1},  for 3-state Potts model \ref{3-Potts1} and 6-state Potts models \ref{6-Potts1} in critical temperatures.  Similar plots were presented in \cite{Jan}. As it is depicted on Figures 2 the power law distribution is complicated by a large fluctuations of  large but rare events which occur in the tail and it appears as noisy curve on the plot. The straight line on  Figure \ref{Fig2}  resents the Pareto distribution.

\begin{figure}[h]
\begin{center}
\subfigure[Ising]{
\includegraphics[width=7cm]{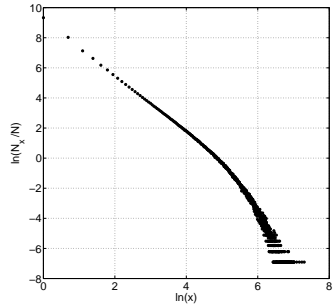}
\label{Ising2}
}

\subfigure[3-Potts]{
\includegraphics[width=6cm]{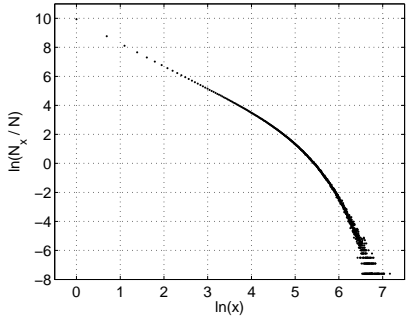}
\label{3-Potts2}
}
\subfigure[6-Potts]{
\includegraphics[width=6cm]{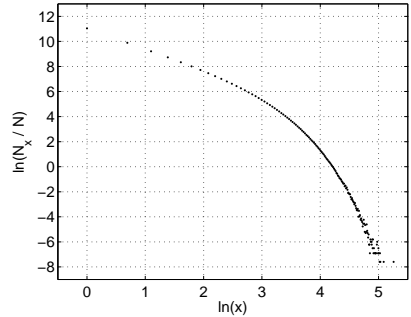}
\label{6-Potts2}
}
\caption{\label{Fig3} The dependence of the  number of domains with mass x normalized by the number of configurations $(N_x/N)$ on the mass(x) in the log-log scale, \ref{Ising2} for the two dimensional Ising model for the  temperature $T= 3.3333 $   , \ref{3-Potts2} for the 3-state Potts model for the  temperature $T= 2.5000  $ and \ref{6-Potts2} for the 6-state Potts model,  for the  temperature $T=2.0000$ .(for 2b and 2c see \cite {Klt})}
\end{center}
\end{figure}

{\bf Figure  \ref{Fig3}} presents the dependence between the number of domains with mass $x$ normalized by number of configurations and mass $x$ in log-log scale \ref{Ising2} for  Ising model for temperature $T=3.3333$ and $L=500$, \ref{3-Potts2} for 3-state Potts model for $T=2.5$ and \ref{6-Potts2}  for 6-state Potts models for $T=2$ and $L=600$.   As it is in Figure 2  the noisy curve appears on the plot. For both models, distributions of domain have a smaler tails than the one for critical temperature,but still bulk of distributions occurs for fairly small sizes. Similar as in the percolation theory \cite{Sta} beyond a critical region the probability of the event with spin in a fixed position belonging to a domain with mass $x$ is asympthotically equal $\rho(x)\sim x^{\tau}\exp[{-\theta x}]$,  where $\theta\sim |T-T_c|^{\frac{1}{\sigma}}$ when $ T \rightarrow T_{c} $.

\begin{figure}[h]
\centerline{\includegraphics [width=9cm]{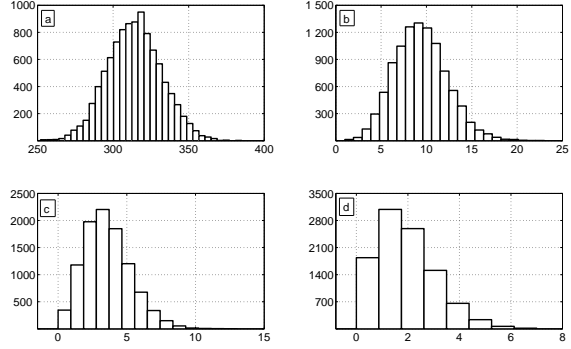}}
\caption{\label{Fig4} Histogram of the number of cluster configurations for fixed mass $m=5$ (a), $m=10$ (b), $m=50$ (c) and $m=70$ (d), for the critical  temperature $T=2.2692$ for the system with $L=600$ and for $10000$ configurations.}
\end{figure}

\begin{figure}[h]
\centerline{\includegraphics[width=9cm]{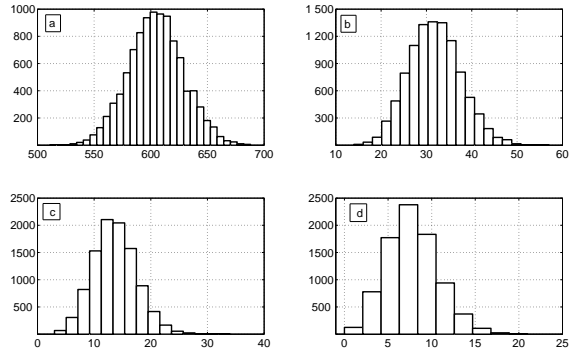}}
\caption{\label{Fig5} Histogram of the number of cluster configurations for fixed mass $ m=5$ (a), $m=10$ (b), $m=50$ (c) and $m=70$ (d), for  the  temperature $T= 4.000$,  for the system with $L=600$, for 10000 configurations.}
\end{figure}

{\bf Figures \ref{Fig4}}  and {\bf\ref{Fig5}} present histograms of  the numbers of cluster configurations with fixed mass for the Ising model. Histograms are presented for masses $x=5$, $x=10$, $x=50$ and $x=70$ for total number of configurations $10000$
of the the system with $L=600$ near critical point and beyond it, respectivelly. For example from histogram $4c$ it can be  seen that five clusters of mass $m=50$ appear  in the $1250$ configurations out of $10000$.

 The test for the gaussian distribution was performed using the Kolmogorov-Smirnov  and the chi-square tests. Near the critical point the gaussian distributions for the number of cluster configurations of a fixed masses are present only for the sizes smaller than 3. Beyond the critical point the gaussian distributions are present only for masses smaller than 5. For example for the temperature $T=4.000$ and the cluster of size 5 the p-value of the chi-squared test was equal to $0.0771$, while for crtical
 temperature and the cluster 5 the p-value of the chi-squared test was equal to 0.0117.\\
 For the clusters of larger size  both for the critical point and outside of it, the histograms of distributions of numbers of domain configurations provide right-skewed distributions.  For  very small masses approximate by the gaussian distributions are present.\\
Experimentally Zipf and  Pareto  laws in the context of phase transitions distributions were investigated in physics on  various occasions,for example: in multifragmantation of atomic nuclear reactions \cite{Ma}, \cite{Ma1}, experimental test for Zipf's law in liquid crystal \cite{Sci}, the discontinous metal films on dielectric substraces \cite{Dob}.
\section{Conclusions}
\label{4}
In our paper we investigate numerically the statistical distributions of domains in Ising and Potts models for critical region as well as beyond it.
The main conclusions are following:
\begin{description}
\item[(i)] In the critical region as well beyond it the log-log plot (Figure 1) of distributions for domain masses presents a strait lines,what means that domains forms a hierarchical order. When $T<T_{c}$ slopes of the lines are bigger  than for critical temperature slope which is  equal to $-1$, when  Zipf's law is valid. Zipf's law  can indicate the presence of phase transition.

\item[(ii)] The distributions of domain masses near the critical point (Figure 2) is well approximated by Pareto distribution, presented in log-log by strait lines. In such case  the phase transition occurs.\\ Beyond the critical region(Figure 3) we do not have strait lines and the tails of the distributions are smaller than the ones described by Pareto tail.

\item[(iii)] Histograms of the numbers of domain configurations with the same number of domains with fixed mass present in the critical point as well as beyond it the right-skewed distributions
\end{description}


\begin{thebibliography}{00}
\bibitem{Zip} Zipf~G.H., Human Behaviour and the Principle of Least Effort, Cambridge, MA: Addison-Wensley, 1949.
\bibitem{Bat} Batty~M., In: Hierarchy in Natural and Social Sciences, Pumain D., (Ed.), Kluwier, Dortrecht, 2005. 1.
\bibitem{Bou} Bouchaud~J.P., In: Workshop on Levy Flights and Related Topics in Physics (Nicea, France 27-30 June 1994), Shlesinger~M.F.,Zas{\l}awsky ~G.M., (Eds.), Berlin-Springer, 1994, 239.
\bibitem{Mbg} Montegna~R.N., Buldyrev~S.V., Goldberger~A.L., Havlin~S., Peng~C.K., Simons~M. and Stanley~H.E.,Phys. Rev. Lett\textbf{73}, 1994, 23.
\bibitem {Sne} Sneppen~K., Zochi~G., Physics in molecular biology, Cambrigde,University press, 2005, 216-218.
\bibitem {Ma} Ma~Y.G., Phys. Rev. Lett \textbf{83} 1999, 3617.
\bibitem {Ma1} Ma~Y.G. et al (NIMROD Collaboration) Phys. Rev. C \textbf{71} ,2005, 054606.
\bibitem{Sci} Sicilia~A., Arenzon~J.J., Dierking~I., Bray~A.J., Cugliandolo~L.F., Martinez-Perdiguero~J., Alonso~I., Pintre~~I.C., Phys. Rev. Lett \textbf{101}, 2008, 19780;
\bibitem{Sta}Stauffer~D,Aharony~A., Introduction to Percolation Theory, Taylor and Francis, London, 1994.
\bibitem{Wat} Watanabe~M.S., Phys. Rev. E \textbf{53}, 1996, 4187.
\bibitem{Car} Cardy~J., Ziff~R.M., J. Stat. Phys. \textbf{110}, 2003, 1.
\bibitem{Jan} Janke~W., Schakel~A.M.J., Phys. Rev. E \textbf{71}, 2005, 708.
\bibitem{Klt} Lukierska-Walasek~K., Topolski~K., Rev. Adv. Matter. Sci., \textbf{23}, 2010, 141.
\bibitem{Klt2} Lukierska-Walasek~K., Topolski K., Comp. Meth. Techn \textbf{16}, 2010, 173.
\bibitem{Fis} Fisher~M.E.,Rep.Prog.Phys.\textbf{30}, 1969, 615.
\bibitem{Man} Mandelbrot~B.B., The Fractal Geometry of the Nature, W.H. Freeman and Company, New York 1983.
\bibitem{Har} Harremoes~P., Topsoe~F., Entropy \textbf{3}, 2001, 191.
\bibitem{Bib} Bibby~B.M. and Sorensen~M., Hiperbolic process in Finance, in Rachev~S.(ed) Handbook of Heavy Tailed Distributions in Finance. Amsterdam, Elseview Science, 2003, 211-248.
\bibitem{Pot} Potts~R.B., Proc. Camb. Phil. Soc \textbf{48}, 1952, 106.
\bibitem{Isi} Ising~E., Z. Physik \textbf{31}, 1925, 253-258.
\bibitem{Wu} Wu~F.Y., Rev. Mod.Phys. \textbf{54} (182), 235.
\bibitem{Fok} Fortuin~C.M., Kasteleyn~P.W., Physica \textbf{57}, 1972, 536-564.
\bibitem{Sww} Swendsen~R., Wang~J., Phys. Rev. Lett. 1986, \textbf{58}, 86.
\bibitem{Tro} Troll~G, Beim Graben~P., Phys.Rev.E \textbf{57}, 1998, 1347.
\bibitem{Dob} Dobierzewska-Mozrzymas~E.,Biega\'nski~P.,Pieciul~E.,W\'ojcik~J., J. Phys. Cond. Matter \textbf{11}, 1999, 5561-68.
 \end{thebibliography}
\end{document}